\pdfoutput=1

\documentclass{tlp}
\usepackage{amsmath}
\usepackage{amssymb}
\usepackage{inlinenum}
\usepackage{helvet}
\usepackage{courier}
\usepackage{xspace}
\usepackage{graphicx}

\newcommand\bcmdtab{\noindent\bgroup\tabcolsep=0pt%
  \begin{tabular}{@{}p{10pc}@{}p{20pc}@{}}}
\newcommand\ecmdtab{\end{tabular}\egroup}

\newcommand{\ignore}[1]{}

\newcommand{\Tr}{\mbox{\bf t}}
\newcommand{\Fa}{\mbox{\bf f}}

\newcommand{\rul}{\leftarrow}
\newcommand{\xxx}{\overline{x}}
\newcommand{\ddd}{\overline{d}}

\newcommand{\lev}[2]{\ensuremath{l_{#1}^{#2}}}
\newcommand{\openset}[2]{BL_{open}(#1, #2)}
\newcommand{\defset}[2]{BL_{def}(#1, #2)}
\newcommand{\gt}{G(T)}
\newcommand{\dlt}{DL(T)}

\newcommand{\id}[1]{\ensuremath{\left \{ \begin{array}{l}#1\end{array} \right \} }}
\newcommand{\lfd}[1]{\ensuremath{\left \lfloor \begin{array}{l}#1\end{array} \right \rfloor }}
\newcommand{\gfd}[1]{\ensuremath{\left \lceil \begin{array}{l}#1\end{array} \right \rceil}}

\newcommand{\D}{D\xspace}
\newcommand{\FD}{\ensuremath{\mathcal{D}}\xspace}
\newcommand{\LFD}{\ensuremath{\Delta}\xspace}
\newcommand{\GFD}{\ensuremath{\nabla}\xspace}

\newcommand{\defp}[1]{\ensuremath{\mbox{\it def}(#1)}\xspace}
\newcommand{\openp}[1]{\ensuremath{\mbox{\it open}(#1)}\xspace}
\newcommand{\Op}[2]{\ensuremath{\Gamma_{#1}^{#2}}}
\newcommand{\lfp}[1]{\ensuremath{\mbox{\it lfp}(#1)}\xspace}
\newcommand{\gfp}[1]{\ensuremath{\mbox{\it gfp}(#1)}\xspace}

\newcommand{\Voc}{\ensuremath{\Sigma}\xspace}
\newcommand{\Rules}{\ensuremath{\mathcal{R}}\xspace}

\newcommand{\res}[2]{{{#1}\rvert_{#2}}}

  \title{FO(FD): Extending Classical Logic with Rule-Based Fixpoint Definitions}

   \author[P. Hou, B. de Cat and M. Denecker]
   {PING HOU, BROES DE CAT and MARC DENECKER\\
         Department of Computer Science, K.U.Leuven, Belgium\\
         \email{\{ping.hou, broes.decat, marc.denecker\}@cs.kuleuven.be}}

\newtheorem{lemma}{Lemma}[section]
\newtheorem{theorem}[lemma]{Theorem}

\newtheorem{corollary}[lemma]{Corollary}
\newtheorem{proposition}[lemma]{Proposition}
\newtheorem{example}[lemma]{Example}
\newtheorem{definition}[lemma]{Definition}

\begin{document}

\maketitle

\begin{abstract}
We introduce fixpoint definitions, a rule-based reformulation of fixpoint constructs. The logic FO(FD), an extension of classical logic with fixpoint definitions, is defined. We illustrate the relation between FO(FD) and FO(ID), which is developed as an integration of two knowledge representation paradigms. The satisfiability problem for FO(FD) is investigated by first reducing FO(FD) to difference logic and then using solvers for difference logic. These reductions are evaluated in the computation of models for FO(FD) theories representing fairness conditions and we provide potential applications of FO(FD).
\end{abstract}

\section{Introduction}
Two mainstream knowledge representation paradigms of the moment are on the one hand, classical logic-based approaches such as description logics \cite{dlog/2003handbook}, and on the other hand, rule-based approaches from logic programming and extensions such as Answer Set Programming and Abductive Logic Programming \cite{baral03,logcom/KakasKT92}. The latter disciplines are rooted in the discipline of Non-Monotonic Reasoning \cite{ai/McCarthy86}. FO(ID)~\cite{tocl/DeneckerT08} integrates both paradigms in a tight, conceptually clean manner. The key to integrate ``rules'' into classical logic (FO) is the observation that natural language, or more precisely, the informal language of mathematicians, has an informal rule-based construct: the construct of {\em inductive/recursive definitions} (IDs).
FO(ID) extends FO not only with an inductive definition construct but also with an expressive and precise non-monotonic reasoning principle. It is an extension of FO with inductive definitions and an integration of FO and LP. It integrates monotonic and non-monotonic logics. The inductive definition construct of FO(ID) formally generalizes Datalog \cite{aw/AbiteboulHV95}. FO(ID) is also strongly related to fixpoint logics.  Monotone definitions in FO(ID) are a different rule-based syntactic sugar of the fixpoint formulas of Least Fixpoint Logic (LFP) \cite{mi/Park70}. Last but not least, FO(ID), being a clear, well-founded integration of rules into classical logic, might play a unifying role in the current attempts of extending FO-based description logics with rules \cite{esws/VennekensD09}. It thus appears that FO(ID) occupies quite a central position in the spectrum of computational and knowledge representation logics.


The work in this paper is inspired by work on FO(ID) to integrate LP-style rules into fixpoint constructs. The resulting constructs are called {\em fixpoint definitions} (FDs). Fixpoint definitions use the rule-based format which will enable us to more easily link fixpoint constructs with the rule-based knowledge representation paradigm and the FO(ID) formalism. We define the logic FO(FD), which is an extension of classical logic with fixpoint definitions. In FO(FD), almost all kinds of inductions can be expressed in a natural way. The study of FO(FD) contributes to the understanding of rule-based systems and thus, to the study of the relation between non-monotonic inductive definitions and fixpoint definitions, to the study of the correspondence between well-founded and fixpoint semantics and to the integration of classical logic-based and rule-based approaches for knowledge representation.

We investigate the connection between FO(FD) and FO(ID) by presenting equivalence preserving transformations from FO(ID) to FO(FD). It turns out that all kinds of inductive definitions in FO(ID) can be expressed in FO(FD). Meanwhile, due to the allowance of the nesting of least and greatest fixpoint constructs in FO(FD), the nesting of induction and coinduction can be represented in FO(FD). Thus, some concepts, e.g., infinite structures and the nesting of recursion and corecursion~\cite{Barwise96}, which can not be defined in FO(ID) in a well-founded way, can be handled naturally in FO(FD). We show that in general, FO(FD) is strictly more expressive than FO(ID).

On the computational level, the satisfiability problem for FO(FD), deciding the satisfiability of FO(FD) theories, is a major research topic. One research direction is towards developing solvers for extensions of propositional logic, e.g., SMT. Difference logic~\cite{cav/NieuwenhuisO05} can be seen as an instance of an SMT framework where propositional logic is extended with simple linear constraints. Efficient implementation techniques for difference logic are emerging in the SMT domain~\cite{cav/NieuwenhuisO05,sat/cotton06}, which makes it a good choice as base technology. In this paper, we develop translations from FO(FD) to difference logic, based on similar reductions of logic programs presented in \cite{lpnmr/JanhunenNS09,amai/Niemela08}. The translations reduce the satisfiability check of FO(FD) theories to finding satisfying interpretations of difference logic theories. This provides a novel approach to model expansion for FO(FD). We also present experimental results.

The paper is organized as follows. In Section~\ref{sec:fd}, we introduce fixpoint definitions and the logic FO(FD). FO(ID) and the relationship between FO(FD) and FO(ID) are presented in Section~\ref{sec:gid}. We investigate the satisfiability problem for FO(FD) by providing the reductions from FO(FD) to difference logic in Section~\ref{sec:satfd}. The reductions are evaluated experimentally in Section~\ref{sec:exper}. In Section~\ref{sec:app}, we present some potential applications of FO(FD) and a conclusion follows in Section~\ref{sec:con}.
\section{FO(FD): A logic of fixpoint definitions}\label{sec:fd}

In this section, we extend first-order logic (FO) with an alternative rule-based fixpoint construct: the construct of {\em fixpoint definitions} (FDs), to formalize a new logic FO(FD), which can be viewed as an extension of first-order logic with mixed induction and coinduction.

\subsection{Syntax}

We assume familiarity with classical logic. A vocabulary $\Voc$ consists of a set of predicate and function symbols.
Terms and FO formulae are defined as usual, and are built inductively from variables, constant and function symbols, logical connectives ($\neg$, $\land$, $\lor$) and quantifiers ($\forall$, $\exists$). Note that predicate symbols occurring in a fixpoint definition are viewed as predicate constants but not predicate variables.

A {\em rule} over a vocabulary $\Sigma$ is an expression of the form $\forall \xxx ( P(\xxx) \rul \varphi[\xxx])$, where $P$ is a predicate symbol of $\Sigma$ and $\varphi[\xxx]$ is an arbitrary first-order formula over $\Sigma$. Atomic formula $P(\xxx)$ is known as the {\em head} of the rule and $\varphi[\xxx]$ is known as the {\em body} of the rule. The {\em defined predicate} of the rule is $P$. The connective $\rul$ is called {\em definitional implication} and is to be distinguished from material implication $\supset$, an abbreviation for $\lnot body \lor head$. We say that a predicate symbol occurs positively (negatively) in a formula if it occurs in the scope of an even (odd) number of negations. A rule is {\em positive} in a set of predicate symbols if these symbols occur only positively in $\varphi$.

For a set $\Rules$ of rules, we denote $\defp{\Rules}$ as the set of defined predicates of its rules, and we denote $\openp{\Rules}$ as the set of all other symbols occurring in $\Rules$.

Without loss of generality, we assume from now on, that rule sets contain for each of its defined predicates exactly one rule of the form $\forall \xxx (P(\xxx) \rul \varphi_P[\xxx])$. Indeed, any set of rules $\{ \forall \bar{x}(P(\bar x) \rul \varphi_1[\xxx]), \ldots, \forall \bar{x}(P(\bar x) \rul \varphi_n[\xxx]) \}$ can be transformed into a single rule $\forall \bar{x}(P(\bar x) \rul \varphi_1[\xxx] \lor \ldots \lor \varphi_n[\xxx])$.

\begin{definition} \label{def:fixpointdefinition}
We define a {\em least fixpoint definition} (LFD), respectively {\em greatest fixpoint definition} (GFD) over vocabulary $\Voc$ by simultaneous induction, as a finite expression $\FD$ of the form $$\lfd{ \Rules, \LFD_1,\dots,\LFD_m, \GFD_1,\dots,\GFD_n} \mbox{, respectively} \gfd{ \Rules, \LFD_1,\dots,\LFD_m, \GFD_1,\dots,\GFD_n}$$ with $0\leq n, m$ such that:
\begin{enumerate}
\item \Rules is a set of rules over \Voc.
\item Each $\LFD_i$ is a least fixpoint definition and each $\GFD_j$ is a greatest fixpoint definition.
\end{enumerate}
To express the remaining conditions, we need some auxiliary concepts and notations. For such an expression $\FD$, we say that a rule $r$ is {\em locally defined} in $\FD$ if $r \in \Rules$, and that a predicate $P$ is {\em locally defined} in $\FD$ if $P\in\defp{\Rules}$, and that $P$ is {\em defined} in $\FD$ if $P$ is locally defined in $\FD$ or defined in any of its subdefinitions $\LFD_1,\dots,\GFD_n$. The set of defined predicates of $\FD$ is denoted $\defp{\FD}$. A symbol is {\em open} in $\FD$ if it occurs in $\FD$ and is not defined in it. The set of open symbols of $\FD$ is denoted $\openp{\FD}$.
\begin{enumerate}
\setcounter{enumi}{2}
\item Every defined symbol of $\FD$ has only positive occurrences in the bodies of rules in $\FD$.
\item Each symbol $P\in\defp{\FD}$ has exactly one local definition in $\FD$. Formally, $\{ \defp{\Rules}, \defp{\LFD_1},\dots,\defp{\GFD_n}\}$ is a partition of $\defp{\FD}.$
\item For every subdefinition $\FD'$ of $\FD$, $\openp{\FD'}\subseteq\openp{\FD}\cup\defp{\Rules}$. In  particular, a symbol defined in another subdefinition $\FD''\neq \FD'$, does not occur in $\FD'$.
\end{enumerate}
A {\em fixpoint definition} is either a least fixpoint definition or a greatest fixpoint definition. We allow arbitrary nesting of least and greatest fixpoint definitions.
\end{definition}

An {\em FO(FD) formula} is either an FO formula or a fixpoint definition. An {\em FO(FD) theory} is a set of fixpoint definitions and FO sentences.

\begin{example}\label{ex:infinitetr}
Assume a binary predicate $T$ denoting a transition graph on a set of vertices, representing the states. Assume a property on states $R$, i.e., a unary predicate on vertices. The set of states $P$ that have an (infinite) path passing an infinite number of times through a state satisfying $R$, is defined by:
\[\gfd{ \forall x\ (P(x) \rul Q(x))\\
			\lfd{ \forall x\ (Q(x) \rul R(x) \land \exists y (T(x,y) \land P(y)))\\
					\forall x\ (Q(x) \rul \exists y (T(x,y) \land Q(y)))}}\]
\end{example}

\subsection{Semantics}

The semantics of FO(FD) is an integration of standard FO semantics with fixpoint semantics of definitions. We start by defining the fixpoint semantics.

Given two disjoint first-order vocabularies $\Sigma$ and $\Sigma'$, a $\Sigma$-interpretation $I$ and a $\Sigma'$-interpretation $I'$, the $\Sigma \cup \Sigma'$-interpretation mapping each element $e$ of $\Sigma$ to $e^{I}$ and each $e \in \Sigma'$ to $e^{I'}$ is denoted by $I+I'$. When $\Sigma' \subseteq \Sigma$, we denote the restriction of a $\Sigma$-interpretation $I$ to the symbols of $\Sigma'$ by $\res{I}{\Sigma'}$. For a $\Sigma$-interpretation $I$ and a tuple of domain elements $\ddd$, we denote by $I[\xxx/\ddd]$ the interpretation that has the same domain as $I$, interprets $\xxx = (x_1, \ldots, x_n)$ by $\ddd = (d_1, \ldots, d_n)$, and coincides with $I$ on all other symbols.

With a set $\Rules$ of rules over $\Sigma$ and a (partial) two-valued $\Voc$-interpretation $I$ interpreting at least all open symbols and no defined symbols, i.e., $\Voc\cap\defp{\Rules}=\emptyset$ and $\openp{\Rules}\subseteq\Voc$, there is a standard way of associating an operator $\Op{I}{\Rules}$ on the set of $\defp{\Rules}$-interpretations with the domain of $I$. For two such interpretations $J, K$, we define $\Op{I}{\Rules}(J)=K$ if for every $\forall \xxx (P(\xxx) \rul \varphi_P[\xxx]) \in \Rules$, $P^K =\{ \ddd | (I+J)[\xxx/\ddd] \models \varphi_P[\xxx]\}$.

If each defined symbol in $\defp{\Rules}$ has only positive occurrences in the body of a rule in $\Rules$, the operator $\Op{I}{\Rules}$ is monotone with respect to the standard truth order on interpretations and hence, it has least and greatest fixpoints in this set denoted $\lfp{\Op{I}{\Rules}}$, respectively $\gfp{\Op{I}{\Rules}}$. Importantly, if $P^I \leq P^{I'}$ for every symbol $P\in\openp{\Rules}$ with only positive occurrences in rule bodies of $\Rules$, then $\lfp{\Op{I}{\Rules}}\leq \lfp{\Op{I'}{\Rules}}$ and $\gfp{\Op{I}{\Rules}}\leq \gfp{\Op{I'}{\Rules}}$.

Given an expression $\FD$ which might be an LFD or a GFD, and an $\openp{\FD}$-interpretation $I$ interpreting at least all open symbols of $\FD$ and no defined ones. We define an operator $\Op{I}{\FD}$ on the set of $\defp{\FD}$-interpretations with domain $dom(I)$. This operator is monotone with respect to the standard truth order on interpretations and hence, it has least and greatest fixpoints in this set. We define $\Op{I}{\FD}(J)$ inductively as the interpretation $K+K'$ where
\begin{itemize}
\item $K$ is the $(\defp{\FD}\setminus\defp{\Rules})$-interpretation
  such that, for $J' =
  I+J|_{\defp{\Rules}}$:
  \begin{itemize}
  \item  $K|_{\defp{\LFD_i}} = \lfp{\Op{J'}{\LFD_i}}$ for all $i= 1, \ldots, m$.
  \item $K|_{\defp{\GFD_j}} = \gfp{\Op{J'}{\GFD_j}}$ for all $j=1, \ldots, n$.
\end{itemize}
Observe that $J'$ interprets all open symbols in every
subdefinition of $\FD$.
\item $K'$ is the $\defp{\Rules}$-interpretation $\Op{I+K}{\Rules}(J|_{\defp{\Rules}})$.
\end{itemize}

\begin{definition}[Model of $\FD$]
Let $\FD$ be a fixpoint definition and $I$ a two-valued $\Voc$-interpretation such
that $\Voc$ contains all symbols in $\FD$. If $\FD$ is an LFD, then $I$ {\em satisfies} $\FD$, or $I$ is a {\em model} of $\FD$, iff $I|_{\defp{\FD}} =
\lfp{\Op{I|_{\openp{\FD}}}{\FD}}$. If $\FD$ is a GFD, then $I$ {\em satisfies} $\FD$, or $I$ is a {\em model} of $\FD$, iff $I|_{\defp{\FD}} = \gfp{\Op{I|_{\openp{\FD}}}{\FD}}$. As usual, this is denoted $I \models \FD$.
\end{definition}

\begin{example}[Continued 2.2]
Semantically, the fixpoint definition in Example 2.2 has the following meaning: the relationship $P$ is the result of iteratively computing a least (for $P$) and a greatest fixpoint (for $Q$). In the $n$-th iteration of the outer fixpoint, $P$ will contain a vertex iff it has a (finite) path that goes through at least $n$ times through vertices with property $R$. At fixpoint, $P$ (and $Q$) will contain a vertex iff it has a path that infinitely often reaches a vertex with property $R$.
\end{example}

\begin{definition}[Model of an FO(FD) theory] \label{def:modelfofdtheory}
Let $T$ be an FO(FD) theory over $\Sigma$ and $I$ a two-valued $\Sigma$-interpretation. Then $I$ is a {\em model} of $T$, denoted by $I \models T$, iff $I \models \varphi$ for every $\varphi \in T$.
\end{definition}

\begin{definition}[Equivalence]
A theory $T_1$ with vocabulary $\Sigma_1$ is equivalent to a theory $T_2$ with vocabulary $\Sigma_2$ iff each model $M_1$ of $T_1$ restricted to $\Sigma_2$ can be extended to a model $M_2$ of $T_2$ and vice versa.
\end{definition}

\subsection{PC(FD)}
In this section, we introduce PC(FD), the propositional fragment of FO(FD). We assume familiarity with propositional logic.

A propositional vocabulary $\Voc$ is a set of propositional atoms. A literal is an atom $p$ or its negation $\neg p$. An atom $p$ is called
a {\em positive} literal, $\neg p$ a negative one. For a literal $l$, we identify $\neg \neg l$ with $l$.


A propositional fixpoint definition is a fixpoint definition such that all symbols occurring in it are propositional symbols.

\begin{example} \label{ex:propositionalfixpoint}
Consider the propositional fixpoint definition
$$\FD =
\lfd{p\rul q \lor r\\
q \rul p \\
\gfd{r \rul p\\
     s \rul t \lor a \\
     t \rul s}}$$
It is obvious that $a$ is the only open atom in this fixpoint definition. There are only two interpretations satisfying $\FD$, namely, $I_1 = \{ a \mapsto \Fa, p \mapsto \Fa, q \mapsto \Fa, r \mapsto \Fa, s \mapsto \Tr, t \mapsto \Tr\}$ and $I_2 = \{ a \mapsto \Tr, p \mapsto \Fa, q \mapsto \Fa, r \mapsto \Fa, s \mapsto \Tr, t \mapsto \Tr \}$. The construction of $I_1$ is illustrated as follows: $I_{1}^{1} = \{ a \mapsto \Fa, p \mapsto \Fa, q \mapsto \Fa, r \mapsto \Tr, s \mapsto \Tr, t \mapsto \Tr \}$ and, because the body of the only rule for $r$ is false, $I_{1}^{2} = \{a \mapsto \Fa, p \mapsto \Fa, q \mapsto \Fa, r \mapsto \Fa, s \mapsto \Tr, t \mapsto \Tr \}$, which is the limit of the iterations and thus, $I_1$ = $I_1^{2}$.
\end{example}

A propositional fixpoint definition $\FD$ is in {\em definitional normal form} (DefNF) if for any $p \in \Voc$, the fixpoint definition contains at most one rule $p \rul \varphi_p$, and either $\varphi_p = \bigvee B_p$ or $\varphi_p = \bigwedge B_p$, where $B_p$ is a set of literals called the {\em body literals}. Any propositional fixpoint definition can be transformed into DefNF in polynomial time using Tseitin transformation~\cite{Tseitin68eng}. Hence without loss of generality, we can from now on assume that propositional fixpoint definitions are in DefNF.

A {\em PC(FD) theory} is a set of propositional formulas and propositional fixpoint definitions. An interpretation $I$ satisfies a PC(FD) theory if it satisfies every formula and every definition of the theory.

\section{A comparison of FO(FD) and FO(ID)}\label{sec:gid}
FO(ID) is an extension of first-order logic with a new construct, namely {\em generalized inductive definitions}, for representing definitions that occur often in mathematics, but in general cannot be expressed in first-order logic. It was originally introduced in~\cite{Denecker:CL2000}, and further developed in~\cite{tocl/DeneckerT08}. In this section, we compare FO(FD) to FO(ID) by providing transformations from generalized inductive definitions to alternating fixpoint definitions and showing that in general, the FO(FD) formalism is strictly more expressive than the FO(ID).

\begin{definition}
Let $\Sigma$ be a vocabulary. A {\em (generalized) inductive definition (GID)} $D$ over $\Sigma$ is a finite set of rules over $\Sigma$. Its sets of defined symbols
$\defp{\D}$, respectively open symbols $\openp{D}$ are defined as usual.
\end{definition}
We do not insist on defined predicates to occur positively in rule bodies in a
generalized inductive definition, but allow non-monotone inductive definitions.

An {\em FO(ID) formula} is a Boolean combination of FO formulas and
generalized inductive definitions. An {\em FO(ID) theory} is a set of
generalized inductive definitions and FO sentences. A model of a generalized inductive definition is a two-valued well-founded model~\cite{tocl/DeneckerT08}. The semantics of FO(ID) is an integration of standard two-valued FO semantics with the well-founded semantics of generalized inductive definitions.

\begin{example}
Consider the following non-monotone inductive definition of even
and odd numbers over the structure of the natural numbers with zero and the successor function:
\[\id{ \forall x (Even(x) \rul x=0 \lor \exists y (x = s(y) \land
\neg Even(y)))\\
\forall x (Odd(x)\rul \exists y (x = s(y) \land Even(y)))}\]
\end{example}

We begin our comparison of FO(FD) and FO(ID) by presenting equivalence preserving transformations from generalized inductive definitions to alternating fixpoint definitions. New symbols may be introduced to the original vocabulary $\Sigma$.

\begin{definition}
Let $D$ be a generalized inductive definition. For each defined predicate $P$ of $D$, we introduce a new predicate symbol $P^{\neg}$ of the same arity of $P$. For each formula $\varphi$, let $\overline{\varphi}$ denote the formula obtained by substituting each negative occurrence $P(\bar t)$ of a defined predicate $P$ in $\varphi$ by $\lnot P^{\neg}(\bar t)$. We define two sets of rules: $\Rules_{D} = \{ \forall \bar{x}(P(\bar x) \rul \overline{\varphi_P[\bar{x}]}) \mid P \in \defp{D} \}$ and $\Rules^{\neg}_{D} = \{ \forall \bar{x}(P^{\neg}(\bar x) \rul \overline{\neg \varphi_P[\bar{x}]}) \mid P \in \defp{D} \}$. Now define $\LFD_{D}$ as $\lfd{  \Rules_\D , \gfd{\Rules^{\neg}_\D}}$.
\end{definition}

Let $D$ be a generalized inductive definition over $\Sigma$. Then $\LFD_{D}$ is a least fixpoint definition over $\Sigma' = \Sigma \cup \{P^{\neg} \mid P \in \defp{D} \}$. Note that $\openp{D} = \openp{\LFD_{D}}$.

\begin{example}[Continued 3.2]
Translating the previous FO(ID) formula into FO(FD) leads to

\[\lfd{ \forall x (Even(x) \rul x = 0 \lor \exists y (x = s(y) \land
{Even}^{\neg}(y))) \\ \forall x (Odd(x) \rul \exists y(x = s(y) \land Even(y))) \\
\gfd{\forall x ({Even}^{\neg}(x) \rul x \not = 0 \land \forall y (x = s(y)
\supset Even(y)))\\ \forall x ({Odd}^{\neg}(x) \rul \forall y (x = s(y) \supset
     {Even}^{\neg}(y)))}}\]
\end{example}

\begin{theorem}
Let $D$ be a generalized inductive definition over $\Sigma$. Then there exists a one-to-one mapping between the $\Sigma$-models $I$ of $D$ and the $\Sigma'$-models $I'$ of $\LFD_\D$ such that the domain of $I$ is the same as that of $I'$, $I' |_\Sigma = I$ and $(P^{\neg})^{I'}$ is the (relative) complement of $P^I$ for each $P \in \defp{D}$.
\end{theorem}

In the following we show that in general, FO(FD) and FO(ID) do not have the same expressive power.

Theorem 4.4 in~\cite{jcss/Schlipf95}, for the well-founded semantics, states that a relation is definable in the well-founded semantics iff it is inductively ($\Pi_{1}^{1}$) definable over the natural numbers. However, on the other hand, Theorem 10 in~\cite{concur/Bradfield96} presents that the FO(FD) alternation hierarchy, the hierarchy of alternating LFD and GFD expressions (ordered along the number of alternations) in any fixpoint definitions, is strict. A consequence is the following result.
\begin{corollary}
FO(ID) is strictly less expressive than FO(FD) on infinite structures.
\end{corollary}

\section{Satisfiability of FO(FD)}\label{sec:satfd}
The second part of this paper presents an approach to \emph{finite model expansion} for FO(FD), the inference task consisting of, given a theory $T$, generating a model for the theory. As a declarative problem solving technique, model generation for FO(FD) allow to represent e.g. temporal properties in an application, increasing its general applicability to among others program verification.

Finite model expansion is equivalent to checking the satisfiability of a Boolean formula, the {\em satisfiability} problem, solved by \emph{SAT solvers}. One approach to check the satisfiability of FO theories, taken by many state-of-the-art solvers, is by reducing the theory to propositional logic (a transformation called \emph{grounding}) and using a SAT solver afterwards. Grounding generally consists of replacing all variables in a formula by all possible substitutions, but intelligent techniques exist that greatly reduce the size of such a grounding, see e.g. \cite{aaai/WittocxMD08}.

Satisfiability checking of FO(FD) theories can be done in a similar way. First the FO(FD) theory is grounded to a PC(FD) theory. Afterwards, the PC(FD) theory is reduced to \emph{difference logic}~\cite{cav/NieuwenhuisO05}, propositional logic extended with linear constraints, and a difference logic solver is used to check the satisfiability of the resulting theory. In the domain of SMT, efficient difference logic solvers have been developed, see e.g. \cite{sat/cotton06}.

\emph{Difference logic}, denoted PC(DL), is the extension of propositional logic with linear difference constraints of the form $x + c < y$, where $x,~y$ and $c$ are integer variables, of which $c$ is known. Syntactically, a linear constraint can occur in the same positions as an atom. An interpretation of a difference logic theory assigns truth values to atoms and integer values to variables.

We first introduce the grounding of FO(FD) to a variable free form. Then, we address the reductions of PC(FD) to difference logic. 

Without loss of generality, we only consider theories in function free FO(FD) for the rest of the paper (any FO(FD) theory can be transformed into a function free theory in polynomial time). 

\subsection{Grounding FO(FD)}
The reduction of an FO(FD) theory $T$ to a PC(FD) theory is defined by:

\begin{definition}
Given an FO(FD) theory $T$ and a finite domain $\mathfrak{D}$. To allow grounding of quantified formulas, we introduce a new constant $c_d$ for each domain element $d \in \mathfrak{D}$, which maps to $d$ in every interpretation $I$. The grounding of $T$ according to domain $\mathfrak{D}$, denoted $\gt$, consists of all $G(\varphi)$ where $\varphi \in T$ and $\varphi$ is either an FO sentence or a fixpoint definition, and $G(\varphi)$ is defined as:
\[
G(\varphi) =
	\begin{cases}
		\bigwedge_{d \in \mathfrak{D}} G(\psi[x/c_d]) & \text{if } \varphi := \forall x \ \psi[x] \\
		\bigvee_{d \in \mathfrak{D}} G(\psi[x/c_d]) 	& \text{if } \varphi := \exists x \ \psi[x]   \\
		G(\psi_1) \wedge G(\psi_2) 		& \text{if } \varphi := \psi_1 \wedge \psi_2\\
		G(\psi_1) \vee G(\psi_2) 		& \text{if } \varphi := \psi_1 \vee \psi_2 \\
		\lfd{G(\psi)} 					& \text{if } \varphi := \lfd{\psi}\\
		\neg G(\psi) 					& \text{if } \varphi := \neg \psi \\
		\gfd{G(\psi)}					& \text{if } \varphi := \gfd{\psi}\\
		p \rul G(\psi)					& \text{if } \varphi := p \rul \psi \text{ and $p$ is an atom} \\
		\psi 							& \text{if } \psi \text{ is an atom} \\
	\end{cases}
\]
\end{definition}

\begin{proposition}
An interpretation $I$ is a model of an FO(FD) theory $T$ iff it is a model of $\gt$.
\end{proposition}

\subsection{Reduction to difference logic}
The aim is to reduce a PC(FD) theory $\gt$ to an \emph{equivalent} theory $\dlt$ in difference logic. The reduction of FO sentences to a PC(DL) theory coincides with their grounding, so for each FO sentence $\varphi \in T$, $\dlt$ contains a sentence $G(\varphi)$. The reduction of fixpoint definitions consists of the \emph{completion} and \emph{level mapping} constraints.

\subsubsection{Completion}
The \emph{completion}, introduced by \cite{adbt/Clark78} for logical rules, expresses in FO the consistency between the truth value of the head and the body of a rule.

The {\em completion} of a propositional rule $r = p \rul \varphi_p$, denoted $Comp(r)$, is given by the formula $p \equiv \varphi_p$. The {\em completion} of a propositional fixpoint definition $\FD$, denoted by $Comp(\FD)$, is $\bigcup_{r \in \FD} Comp(r)$.

An important property is that $I \models \FD$ implies $I \models Comp(\FD)$. The converse is not true, $\FD$ generally has fewer models than $Comp(\FD)$.

\begin{example}\label{ex:completionvsmodel}
Consider the propositional fixpoint definition 
$$\FD = \lfd{ p \rul p \lor a \\
           \gfd{ q \rul q \land p}}$$
Then $Comp(\FD) = (p \equiv p \lor a) \land (q \equiv q \land p)$.
$\FD$ has two models: $\{ a \mapsto \Fa, p \mapsto \Fa, q \mapsto
\Fa \}$ and $\{ a \mapsto \Tr, p \mapsto \Tr, q \mapsto \Tr \}$;
$Comp(\FD)$ has the same two models, and the additional three
models: $\{ a \mapsto \Fa, p \mapsto \Tr, q \mapsto \Tr\}$, $\{ a
\mapsto \Fa, p \mapsto \Tr, q \mapsto \Fa\}$ and $\{ a \mapsto \Tr,
p \mapsto \Tr, q \mapsto \Fa \}$.
\end{example}

\subsubsection{Level mappings}
To obtain equivalence of $T$ and $\dlt$, it is necessary to ensure that only interpretations consistent with the operator $\Op{I}{\FD}$ are models of $\dlt$. We take a \emph{level mapping} approach to characterize the models of the fixpoint operator. This is an extension of the technique presented in \cite{lpnmr/JanhunenNS09,amai/Niemela08}, where stable model generation of logic programs is obtained by reduction to difference logic.

\begin{definition}[level mapping]
Given a fixpoint definition $\FD$, define a function $l_{\FD}: \defp{\FD} \rightarrow \mathbb{N}$, with $\defp{\FD}$ the set of all defined atoms in $\FD$. Function $l$ is then the \emph{level mapping} function and $l_{\FD}(p)$ is the \emph{level} of defined atom $p$ for fixpoint definition $\FD$.
\end{definition}

A level mapping function $l_{\FD}$ is introduced for each (nested) fixpoint definition $\FD$ in $\gt$. In ground form, for each fixpoint definition $\FD$ and for each defined atom $p$ in $\FD$, we introduce an integer variable, denoted $\lev{\FD}{p}$.

The level mapping should ensure that the truth of a least fixpoint relation or the falsity of a greatest fixpoint relation can always be \emph{finitely justified} in terms of locally defined atoms or open ones.

\subsubsection{Level mapping constraints}
We introduce PC(DL) formulas which, as part of $\dlt$, act as constraints on the relation between the levels of different defined atoms within one fixpoint definition. Theory $\dlt$ will be satisfiable iff such a finite justification exists.

As mentioned earlier, all rules are considered to be in DefNF. For a given rule $r$ in fixpoint definition $\FD$, $h$ denotes the head and $body(r)$ is the set of all literals occurring in the body of $r$. The sets $\defset{\FD}{r}$ and $\openset{\FD}{r}$ denote the set of defined, respectively open body literals(BL)
\begin{align}
\defset{\FD}{r} &= \{d | d \in \defp{\FD} \cup \lnot \defp{\FD} \ \text{and} \ d \in body(r)\} \\
\openset{\FD}{r} &= \{o | o \in \openp{\FD} \cup \lnot \openp{\FD} \ \text{and} \ o \in body(r)\}
\end{align}
We now introduce the constraints.

No justification is necessary for an atom defined in a {\em GFD} if it is true, nor for an atom defined in an {\em LFD} which is false. Formally represented by the constraints:
\begin{align}
	\text{if \FD is a GFD: }&\qquad a \supset \lev{\FD}{h}=0\\
	\text{if \FD is an LFD: }& \quad \lnot a \supset \lev{\FD}{h}=0
\end{align}

When an atom defined in a {\em GFD} is not true or an atom defined in an {\em LFD} is not false, a \emph{justification} is necessary. A justification is a set of body literals of a rule sufficient to derive the head in a given interpretation. Although looping is allowed over literals defined in lower fixpoints, it has to be possible to construct a justification which does not loop over literals in the same level.

Deriving that the head of a rule with a disjunctive body in an LFD is true requires only one body atom to be true. If it were a rule with a conjunctive body, all body literals would be necessary as justification. This also holds for the relation between their levels: in the disjunctive rule, the minimal level of all true body literals can act as the level of the justification. In the conjunctive case, the level is the maximum level of all body literals.

These ideas can be generalized and formalized as constraints. For clarity, the constraints are not in PC(DL), but we introduce $min\{\}$ and $max\{\}$ notation to represent respectively the minimum and maximum of a set of levels. Assume an interpretation $I$ to further simplify the aggregate notation. All aggregates can be translated out easily, independent of $I$ (see further). Also assume a fixpoint definition $\FD$ with a locally defined atom $h$ in a rule $r$.
\begin{enumerate}
\item If $\FD$ is an LFD and $r$ has a conjunctive body, the translation of $r$ is:
\begin{equation}\label{eq:maplfdconj}
h \supset \lev{\FD}{h}>max\{\lev{\FD}{d} | d \in \defset{\FD}{r} \ \text{and} \ I(d) =\Tr \}
\end{equation}
\item If $\FD$ is an LFD and $r$ has a disjunctive body, the translation of $r$ is:
\begin{equation}
\begin{split}
h \supset &(\lev{\FD}{h}>min\{\lev{\FD}{d} | d \in \defset{\FD}{r} \ \text{and} \ I(d) = \Tr\} \\ &\lor \bigvee_{d \in \defset{\FD}{r}} d \ \lor \bigvee_{o \in \openset{\FD}{r}} o)
\end{split}
\end{equation}
\item If $\FD$ is a GFD and $r$ has a disjunctive body, the translation of $r$ is:
\begin{equation}
\lnot h \supset \lev{\FD}{h}>max\{\lev{\FD}{d} | d \in \defset{\FD}{r} \ \text{and} \ I(d) = \Fa \}
\end{equation}
\item If $\FD$ is a GFD and $r$ has a conjunctive body, the translation of $r$ is:
\begin{equation}
\begin{split}
\lnot h \supset& (\lev{\FD}{h}>min\{\lev{\FD}{d} | d \in \defset{\FD}{r} \ \text{and} \ I(d) = \Fa \} \\ &\lor \bigvee_{d \in \defset{\FD}{r}} \lnot d \ \lor \bigvee_{o \in \openset{\FD}{r}} \lnot o)
\end{split}
\end{equation}
\end{enumerate}

Similar constraints apply for the level of the head $h$ of rules defined in a subdefinition of $\FD$, but the inequality $\lev{\FD}{h} > \ldots$ is relaxed to $\lev{\FD}{h} \geq \ldots$

\begin{proposition}
The truth value of a higher defined atom can only be justified by finite looping over literals in the same definition or infinite looping over literals in lower definitions. This is expressed by using similar constraints for locally defined rules and for rules defined in subdefinitions, but dropping the strict order requirement on the second, effectively allowing infinite looping over literals defined in subdefinitions.
\end{proposition}

\begin{example}
In the following fixpoint definition, using only strict ordering would lead to a contradiction, although a model exists.
\[\lfd{
	a \rul c \\
	\gfd{
		c \rul d \\
		d \rul c
	}
}\]
\end{example}

\begin{theorem}
If an FO(FD) theory is transformed using the presented reduction to PC(DL) via PC(FD), the resulting PC(DL) theory will be satisfiable iff the FO(FD) theory is satisfiable. Any model of the PC(DL) theory can be transformed into a model of the FO(FD) theory. 
\end{theorem}

\subsubsection{Aggregate reduction}
To obtain PC(DL) constraints, the aggregates $min$ and $max$ have to be transformed into difference constraints, which can be done in the following fashion:

\[
\begin{array}{ll}
\text{Replace }&\lev{\FD}{h}>max(\{\lev{\FD}{d}|d \in \defset{\FD}{r} \ \text{and} \ I(d) =\Tr\}) \\
\text{by }& \bigwedge_{d \in \defset{\FD}{r}} (\lev{\FD}{h}>\lev{\FD}{d} \lor \lnot d) \\
& \\
\text{Replace }&\lev{\FD}{h}>min(\{\lev{\FD}{d}|d \in \defset{\FD}{r} \ \text{and} \ I(d)=\Tr\}) \\
\text{by }& \bigvee_{d \in \defset{\FD}{r}} (\lev{\FD}{h}>\lev{\FD}{d} \land d)
\end{array}
\]

For a condition $I(d)=\Fa$ instead of $I(d) =\Tr$, the literal $d$ is replaced with $\lnot d$.

\subsubsection{Optimization: partial level mapping}
Level mappings constraints are used to enforece dependencies between defined atoms. Often, a preprocessing step (before PC(DL) reduction) allows to deduce that certain atoms will never depend on each other. In that case, less mapping constraints are necessary. A simple example are non-cyclic dependencies, for which no level mapping constraints are necessary ($Comp(\FD)\models \FD$). These dependencies can be obtained by calculating the \emph{strongly connected components} \cite{siamcomp/Tarjan72} on the \emph{dependency graph} of the fixpoint definition, a general technique used among others in stable model generation \cite{lpnmr/SyrjanenN01}.

The \emph{dependency graph} consists of all edges $h \leadsto b$, for each rule $r$ in $\FD$ with head $h$ and for each body literal $b$ of $r$ that is defined in $\FD$ or in a parent of $\FD$. A \emph{strongly connected component} of a directed graph is a maximal subset in which a path exists between any two nodes in the set.

\begin{proposition}
Only defined atoms that are in a strongly connected component with $\|nodes\|\geq 2$ or have recursion over themselves (e.g. $h \leadsto h$) need a level mapping. Body atoms that are not in the same strongly connected component as the head can be treated as open instead of defined atoms.
\end{proposition}

To implement this idea, the set of open body literals $\openset{\FD}{r}$ is redefined: for a rule $r$, a body literal of $r$ is considered open if it is not defined, defined in an ancestor of the definition of $r$ or if it is not in the same strongly connected component as the head of $r$. The set $\defset{\FD}{r}$ contains all remaining body literals.

\subsubsection{Optimization: stronger constraints}
The presented constraints are \emph{weak}: infinitely many models of the PC(DL) reduction exist that are equivalent (modulo shared vocabulary) to one model of the FO(FD) theory. Exact one-to-one mapping is not possible because expressions of the form $x=c$, where $c$ is a known integer constant, cannot be expressed in difference logic. By expressing all constraints in terms of one integer variable, which acts as a floating ground, we can greatly reduce the number of redundant models.

The presented constraints can be adapted to obtain such \emph{stronger} constraints by enforcing that the level of the head of a rule is the minimum allowed by its associated constraint, adapted from in \cite{lpnmr/JanhunenNS09,amai/Niemela08}. For example for a rule with a conjunctive body in a least fixpoint, which is subject to the constraint expressed by equation \ref{eq:maplfdconj}, a second constraint is added of the form:
\begin{equation}
h \supset \bigvee_{d \in \defset{\FD}{r}} (\lev{\FD}{h}= \lev{\FD}{d}+1 \land d)
\end{equation}

\section{Implementation and experiments}\label{sec:exper}
In this section, we report our first experiments, on model checking of fairness conditions, with a prototype implementation of the reductions from FO(FD) to difference logic. We used the $\mu$-calculus fairness expression presented in \cite{tacas/LiuRS98}:
\begin{equation}
\nu X. \mu Y.[-](\langle a \rangle X) \lor Y
\end{equation}
It expresses that a state in the transition system is fair if on all possible paths, an $a$-labeled edge is infinitely often taken. Translated into an FO(FD) theory:
\[
\gfd{ \forall x\ (P(x) \rul Q(x))\\
	\lfd{ \forall x\ (Q(x) \rul \forall y \ (Edge(x, y) \supset (L(y, a) \land P(y)) \lor Q(y)))
	}
}
\]
where the relations $P$ and $Q$ contain states from which infinitely often a state labelled $a$ will be reached. The predicate $L$ is the labelling relation, expressing that a state has a certain label. The predicate $Edge$ is the transition relation.

\begin{figure}
	\centering
	\includegraphics[width=0.7\textwidth]{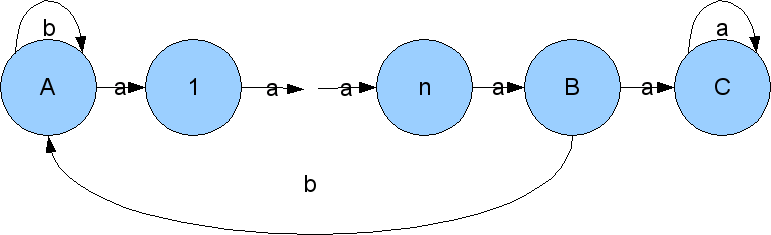}
	\caption{A transition graph}
	\label{fig:transitiongraph}
\end{figure}

The task consists of doing model expansion, where the transitions and labellings are known, to decide which nodes are fair. Both weak and strong constraints were tested. The experiments were done on the graph depicted in Figure~\ref{fig:transitiongraph}. The results of these experiments are as shown in Table~\ref{fig:results}, grounding times are included. The machine used is a dual-core 2.4 GHz with 4 Gb RAM, with Ubuntu 8.04 OS. Yices2 was used as difference logic solver.

\begin{table}
\begin{tabular}{|c|c|c|}
\hline
$\|nodes\|$ & weak(sec) & strong(sec) \\
\hline
503 & 0.011	& 0.004 \\
1503 & 0.21 & 0.09 \\
2503 & 20.51 & 14.19 \\
\hline
\end{tabular}
\caption{Model checking results} \label{fig:results}
\end{table}

From these preliminary results, we conclude that fairness conditions can be evaluated efficiently using our reduction to difference logic. Strong constraints are significantly faster due to their fewer degrees of freedom, which presumably allow more propagation and pruning of the search space. In \cite{wlp/KeinanenN04}, similar results were obtained with the same experiment.

\section{Applications}\label{sec:app}
Many applications can be found on the use of fixpoint expressions. Most of them focus on inductive and coinductive definitions (which have nesting depth 1), used e.g. for expressing transitive closure (reachability), bisimulation and situation calculus. One important application domain for nested fixpoint definitions is the verification of automata. Temporal logics like CTL* allow to express time-variant properties of automata, e.g. fairness. The $\mu$-calculus, a superset logic of those temporal logics bound on fixpoint expressions, can be transformed into fixpoint definitions. So any application of model checking or model generation of temporal logics can be expressed in FO(FD). Another application domain are so-called \emph{parity games}, which are infinite games played on a graph with priority-annotated nodes. For more information we refer to \cite{atva/FriedmannL09}. Parity games can be expressed in fixpoint logic, the nesting increasing polynomially with the number of priorities.

\section{Conclusions and related work}\label{sec:con}
In this paper, we have introduced fixpoint definitions, an alternative rule-based expression of fixpoint constructs, and the logic FO(FD), which is an extension of classical logic with fixpoint definitions. We have compared FO(FD) and FO(ID) by providing equivalence preserving transformations of non-monotone inductive definitions to alternating fixpoint definitions and showed that FO(FD) is strictly more expressive than FO(ID) on infinite structures. We have investigated the satisfiability problem for FO(FD) by developing reductions from FO(FD) to difference logic. Hence, SMT solvers supporting difference logic can be used for computing fixpoint models of FO(FD) theories {\em without any modifications}. We have implemented these reductions and evaluated the resulting solver in the computation of models of FO(FD) theories. In general, our transformation to difference logic is exponential in the nesting depth of a fixpoint definition, but for most practical applications they prove compact and efficient.

$\mu\text{MALL}^{=}$, which is the logic obtained by extending MALL (multiplicative, additive linear logic) with equality, quantification (via $\forall$ and $\exists$) and mixed least and greatest fixpoint constructors, was introduced in~\cite{lpar/BaeldeM07}. It seems that $\mu\text{MALL}^{=}$ has the same expressive power as FO(FD). However, $\mu\text{MALL}^{=}$ is developed from a proof theory standpoint whereas FO(FD) is developed from a model theory point of view.

Gupta et al. in~\cite{iclp/GuptaBMSM07} introduced coinduction, corresponding to the greatest fixpoint constructor, into logic programming to obtain coinductive logic programming. Discussed applications are  verification, model checking, non-monotonic reasoning, etc. However, in coinductive logic programming, naively mixing coinduction and induction leads to contradictions while arbitrary cyclical nesting of least and greatest fixpoint constructs is allowed in FO(FD). Another difference is on the computational level. The main computational task for FO(FD) is model generation in the context of a finite domain. However, model generation in coinductive logic programming is applied to constructs of an infinite Herbrand model based on an infinite Herbrand universe.

Niemel{\"a}, Janhunen et al. in~\cite{lpnmr/JanhunenNS09,amai/Niemela08} introduced stable model generation of general logic programs via reductions to difference logic. They also used stable model generation to find solutions to Boolean equation systems \cite{wlp/KeinanenN04}. This is a related fixpoint formalism, in which among others $\mu$-calculus can be expressed.

There are several other solvers for solving the satisfiability and validity problems for fixpoint logics, e.g.,~\cite{atva/FriedmannL09}. Our reduction is based on SMT solver technology, whereas referenced works are based on characterizations of satisfiability through infinite (cyclic) tableaux. Well-foundedness for unfoldings of least fixpoints is then checked using deterministic parity automata.

\bibliographystyle{acmtrans}
\bibliography{krrlib}

\clearpage

\end{document}